\documentclass[journal]{IEEEtran}
\usepackage{cite}
\usepackage{textcomp}
\usepackage[pdftex]{graphicx}
\usepackage{amsmath,amssymb,amsthm,mathrsfs,amsfonts,dsfont}
\usepackage{indentfirst}
\usepackage{threeparttable}
\usepackage{booktabs}
\usepackage{algorithm,algorithmic}
\algsetup{linenosize=\small}
\usepackage{multirow}
\usepackage{booktabs}
\usepackage{bm}
\usepackage{enumerate}
\usepackage{cite}
\usepackage{epstopdf}
\usepackage{textcomp}
\usepackage{csquotes}
\usepackage{url}
\usepackage{lipsum}
\usepackage{mathtools}
\usepackage{cuted}
\usepackage{xcolor}
\usepackage{array}

\newcommand{\specialcell}[2][c]{%
  \begin{tabular}[#1]{@{}l@{}}#2\end{tabular}}

\ifCLASSINFOpdf
\else
\fi

\def\BibTeX{{\rm B\kern-.05em{\sc i\kern-.025em b}\kern-.08em
    T\kern-.1667em\lower.7ex\hbox{E}\kern-.125emX}}
\begin{document}

\title{V2X Content Distribution Based on Batched Network Coding with Distributed Scheduling}
\author{\IEEEauthorblockN{Yumeng~Gao, \emph{Student Member, IEEE}, Xiaoli~Xu, Yong~Liang~Guan, \emph{Senior Member, IEEE}, and~Peter~Han~Joo~Chong, \emph{Member, IEEE}}%
\thanks
{Y. Gao, and Y. L. Guan are with the School of Electrical and Electronic Engineering, Nanyang Technological University, Singapore 639798 (email: ygao005@e.ntu.edu.sg; eylguan@ntu.edu.sg).}
\thanks{X. Xu is with the School of Electrical and Information Engineering, The University of Sydney, Sydney, NSW 2006, Australia
 (email:xiaoli.xu@sydney.edu.au).}
\thanks{P. H. J. Chong is with the Department of Electrical and Electronic Engineering Auckland University of Technology, Auckland 1142, New Zealand (email: peter.chong@aut.ac.nz).}
\thanks{This work was supported by the NTU-NXP Intelligent Transport System
Test-Bed Living Lab Fund S15-1105-RF-LLF from the Economic Development Board, Singapore.}}
\maketitle
\begin{abstract}
Content distribution is an application in intelligent transportation system to assist vehicles in acquiring information such as digital maps and entertainment materials. In this paper, we consider content distribution from a single roadside infrastructure unit to a group of vehicles passing by it. To combat the short connection time and the lossy channel quality, the downloaded contents need to be further shared among vehicles after the initial broadcasting phase. To this end, we propose a joint infrastructure-to-vehicle (I2V) and vehicle-to-vehicle (V2V) communication scheme based on batched sparse (BATS) coding to minimize the traffic overhead and reduce the total transmission delay. In the I2V phase, the roadside unit (RSU) encodes the original large-size file into a number of batches in a rateless manner, each containing a fixed number of coded packets, and sequentially broadcasts them during the I2V connection time. In the V2V phase, vehicles perform the network coded cooperative sharing by re-encoding the received packets. We propose a utility-based distributed algorithm to efficiently schedule the V2V cooperative transmissions, hence reducing the transmission delay. A closed-form expression for the expected rank distribution of the proposed content distribution scheme is derived, which is used to design the optimal BATS code. The performance of the proposed content distribution scheme is evaluated by extensive simulations that consider multi-lane road and realistic vehicular traffic settings, and shown to significantly outperform the existing content distribution protocols.
\end{abstract}

\begin{IEEEkeywords}
Vehicular networks, joint I2V and V2V communication, content distribution, batched sparse code, channel rank distribution.
\end{IEEEkeywords}


\maketitle
\section{Introduction}\label{sec:introduction}
\IEEEPARstart{V}{ehicular} ad-hoc network (VANET) \cite{hartenstein2009vanet} is an emerging technology that integrates the concept of wireless ad-hoc network and new advances in communication networks to the domain of vehicles. As a key part of intelligent transportation system (ITS), VANET is envisioned and designed to provide a large number of various and attractive applications for road safety concern, navigation, traffic efficiency and commercial interests, etc. Safety-related services such as collision avoidance, emergency warning and blind crossing are the initial motivations to establish ITS \cite{ITS}. With the increasing popularity of vehicular communication system, the designs of non-safety protocols to offer drivers on-board comfort and entertainment extend the mission of ITS. Content distribution or content downloading is an important form of communication to achieve both safety and non-safety types of services. It can be utilized for delivering informative contents like digital maps, traffic information like accident reports, or commercial materials like promotional videos.

Generally, content is distributed by the roadside unit (RSU) to the vehicles which are equipped with on-board units (OBUs) via the infrastructure-to-vehicle (I2V) communications. An explicit requirement of the content distribution protocol is to ensure complete delivery and low downloading delay for all the intended receiving vehicles. However, due to the dynamic vehicular environments and high mobility of vehicles, the channel between the RSU and passing vehicles is prone to packet loss. Meanwhile, considering the cost and location restriction of RSU deployment, in some situations, it is not always possible to have unbroken RSU coverage in certain regions. So the insufficient connection time between the vehicle and the RSU poses challenges on the reception of the complete file. To tackle these problems, the content dissemination process is extended to involve the assistance of vehicle-to-vehicle (V2V) communication. After moving out of the RSU range, vehicles share the collected, possibly incomplete, data with neighboring vehicles and help each other recover the uncollected pieces, performing V2V collaborative sharing. To achieve full downloading with low downloading delay for all vehicles without flooding the V2V channel, it is crucial to have a smart transmission coordination among the vehicles.

Existing content distribution protocols have considered using the routing-based scheduling to determine the V2V sharing phase \cite{lee2002demand,nandan2005co,lee2007first} while network coding (NC)  \cite{ahlswede2000network}\cite{ho2006random} is adopted in \cite{lee2006code,ahmed2006vanetcode,firooz2012collaborative,ye2012efficient,zhu2015multiple,xing2017stochastic,li2011codeon,liu2016network} to enhance the transmission reliability under the fading channel for both I2V and V2V communications. On the basis of NC, \cite{li2011codeon,yu2012rank,liu2016network} propose different scheduling schemes. Nevertheless, they relay on the frequent exchange of the packet reception status and require the successful decoding of every network-coded block at all vehicles, which increases the network overhead and the transmission delay.

Distinct from the above protocols, we propose a joint design for I2V and V2V content distribution based on a two-stage version of the BATS network code \cite{yang2014batched} to improve the effectiveness of the content distribution, achieve low downloading delay, and reduce the transmission overhead. To the best of our knowledge, this is the first work that applies the BATS code to solve the content distribution problem in the V2X network. We consider the scenario where a RSU is installed at the side of the road\footnote{The proposed joint I2V and V2V content distribution scheme can be easily extended to muli-RSU case by considering prolonged connection time.}. The content distribution is completed in two phases: the RSU broadcasting phase and the V2V cooperative sharing phase. In the RSU broadcasting phase, the file is encoded with a rateless code (BATS outer code) at the RSU and the coded batches are sent out sequentially. The rateless nature of the outer code allows the RSU to broadcast continuously, hence no packet reception feedback from the vehicle to the RSU is required and the erasure recovery of each coded packet is enhanced.
After the vehicles leave the coverage area of the RSU, the cooperative sharing phase starts and the received packets are re-encoded with random linear network code (BATS inner code), then transmitted. This V2V transmission should not be uncontrolled, otherwise the V2V channel will be flooded, with no meaningful outcome. We propose a distributed scheduling scheme to prioritize the transmissions with higher utility and finally achieve a short transmission delay. Using the belief propagation (BP) decoding algorithm\cite{yang2014batched}, vehicles are able to decode the distributed content without the need to decode each coded block.

Note that our proposed design can be applied to both DSRC \cite{jiang2006design} and Cellular V2X (C-V2X) \cite{CV2X} technologies. The BATS code is still applicable when considering the cellular base station as the source of content.
The major contributions of this paper are as follows.
\begin{enumerate}[{1)}]
\item
It is new to apply BATS code to address the content distribution problem in the vehicular network. Based on BATS code, we have proposed a joint I2V and V2V distribution protocol and a utility-based distributed V2V scheduling scheme where no packet reception status is exchanged and an individual vehicle can prioritize the transmission sequence in a distributed manner. Compared with the existing protocols with centralized scheduling and frequent status exchange \cite{li2011codeon,zeng2017channel}, our proposed scheme significantly reduces the total transmission delay and the network overhead. The effectiveness and scalability of the proposed scheme are verified through extensive simulations. The total transmission delay is proved to be close to the theoretical lower bound. With the expansion of the size of the vehicle group, the transmission delay can be further reduced.
\item
To optimize the BATS code for achieving high coding efficiency, it is critical to design an optimal degree distribution of BATS code at the source node which is obtained by analyzing the channel rank distribution. Different from the line network topology of the existing work \cite{yang2014batched,ng2013finite}, the cyclic topology and the utility-based scheduling in our studied scenario highly complicate the rank distribution. 
We address this problem by proposing an analytical approach to approximate the rank distribution of all the batches in a simplified but accurate way.
\end{enumerate}
The rest of the paper is organized as follows. Section \ref{related_work} outlines the related work. Section \ref{system} introduces the system model. The first and second phases of the proposed schemes are given in Section \ref{section_P1} and Section \ref{section_P2}. The estimation of rank distribution is in Section \ref{section_rank}. Section \ref{section_simulation} evaluates the performances of the proposed scheme and verifies the effectiveness and superiority of the proposed solution through simulations. Finally, we conclude the paper in Section \ref{section_conclusion}.
\vspace{-0.0cm}
\section{Related Work}\label{related_work}
\subsection{Existing Content Distribution Protocols}
Routing-based V2V sharing protocols for content distribution in vehicular networks were investigated in \cite{lee2002demand,nandan2005co,lee2007first}. However, these protocols are prone to generating duplicate transmissions and significant overhead which prolong the distribution delay. To enhance the transmissions reliability under the fading channel, many existing protocols applied network coding to the content distribution \cite{lee2006code,ahmed2006vanetcode,firooz2012collaborative,ye2012efficient,zhu2015multiple,xing2017stochastic,li2011codeon,liu2016network}. CodeTorrent \cite{lee2006code} was the pioneer work that combined the file swarming with network coding. Some similar schemes \cite{ahmed2006vanetcode,firooz2012collaborative,ye2012efficient} have been conducted to investigate the benefit of the network coded transmissions in the aspect of reducing total downloading delay. W. Zhu \textit{et al.} \cite{zhu2015multiple} extended the work in \cite{firooz2012collaborative} to study the PMFs of the data downloading time for multiple-vehicles by using the random, feedback and NC-based schemes under the perfect channel condition. W. Xing \textit{et al.} \cite{xing2017stochastic} solved the successful transmission probability of the NC-source data when vehicles faced the fading and co-channel interference by using the stochastic geometry.

Unlike the aforementioned schemes which focus on the system analysis but lack careful scheduling design for the V2V sharing, CodeOn \cite{li2011codeon} exploited the symbol-level NC and the exchanged packet decoding status to determine the V2V transmission priority. However, the frequent status exchange leads to large traffic overhead and the symbol-level network coding requires large computational complexity. A SNR-based scheduling strategy was proposed in \cite{zeng2017channel} for data dissemination where large-scale channel loss and vehicle positions were assumed to be well predicted by RLS. However, the impact of the small-scale fading on the utility of the vehicles were not considered which may make difference to the scheduling strategy. Moreover, a centralized TDMA-based scheduling scheme was proposed in \cite{zhang2015novel} to control the V2V communication. K. Liu \textit{et al.} \cite{liu2016network} also utilized the RSU to perform the scheduling decision and built an effective cache strategy but the network overhead due to the frequent packet status exchange could not be avoidable.

To further improve the current content distribution protocols and solve the above mentioned problems, our proposed joint I2V and V2V scheme makes the best of the rateless nature of the BATS code to reinforce the usefulness of each transmitted packet in the lossy channel and introduces a smart distributed scheduling method so that each vehicle can determine the transmission sequence locally.
\subsection{Batched Sparse Code}
BATS code was originally proposed for multi-hop line network in \cite{yang2014batched} and it consists of an outer code and an inner code. The outer code generalizes the fountain code by generating $M$ coded packets from a group of input packets to form a batch, where $M$ is referred to as the batch size.
The BATS inner code which is random linear network code (RLNC), can be performed within each batch at the intermediate nodes such as vehicles in this context, to generate coded packets for V2V transmissions. The outer and inner code can be jointly decoded with efficient belief propagation (BP) algorithm \cite{yang2014batched}.
Compared with the conventional RLNC, BATS code has much lower coding overhead that is determined by the batch size, and lower encoding and decoding complexity, which makes it suitable for the computation-limited devices, e.g., the OBU. Based on the notations in Table \ref{Notation_table}, the encoding complexity of the BATS code is $O(M\ell)$ and the decoding complexity is $O(M^2+M\ell)$ per packet. In contrast, when using the RLNC, the encoding and decoding complexities are $O(F\ell)$ and $O(F^2+F\ell)$, respectively. Since $F\gg M$, BATS code has a much lower encoding/decoding complexity.

The application of BATS code for distributing a file from a common source node to a group of closely-located receivers in a static ad-hoc network has been discussed in \cite{xu2016two}. Our work extends the application of BATS code to V2X networks where each vehicle has a short connection time with the RSU. Different from the static network considered in \cite{xu2016two} where the number of source transmissions can be simply adjusted according to the channel statistics, the broadcasting time of the RSU is determined by the vehicle speed. Moreover, vehicles face a varying channel in the I2V communication. The channel-induced loss is different for each vehicle at different time instances. So it is necessary to investigate the system performance as well as the design of an efficient BATS code.
\vspace{-0.0cm}
\section{Problem Formulation}\label{system}
\subsection{System Model}
\begin{figure*}
\centering
\includegraphics[width=0.72\textwidth]{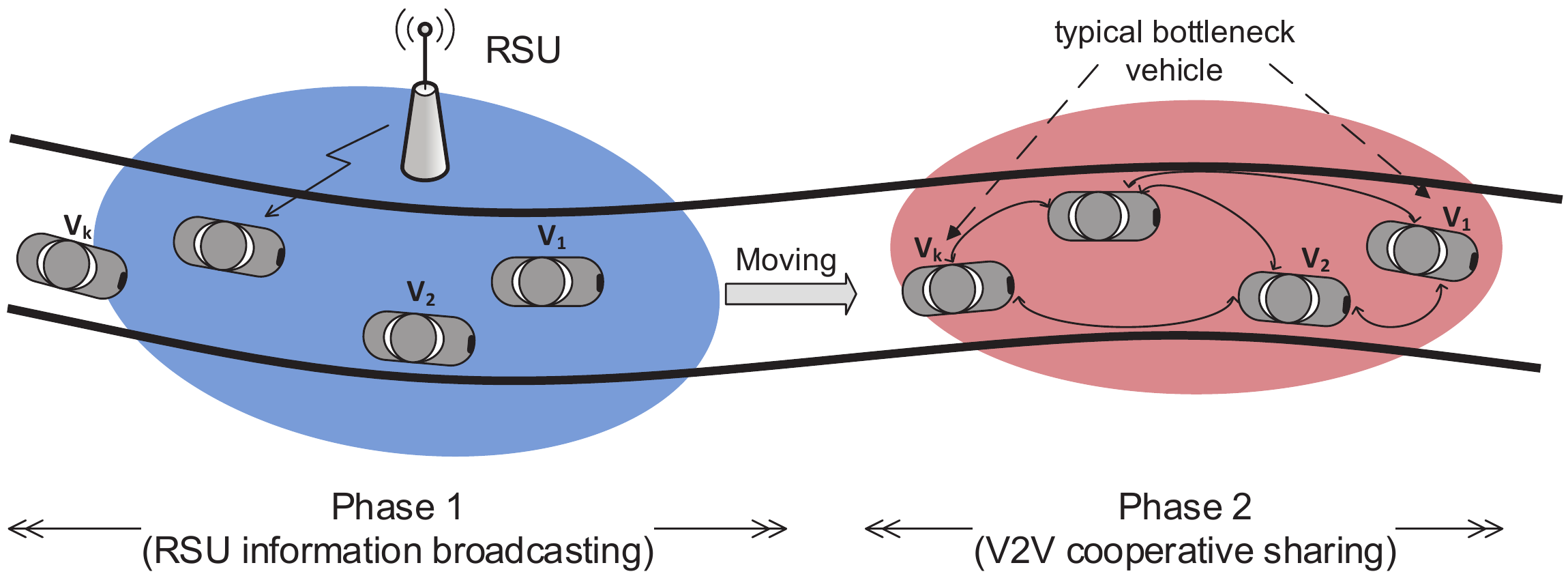}
\caption{The system model of the content distribution.}
\label{system_model}\vspace{-0.5cm}
\end{figure*}

We consider a general system model of content distribution in the vehicular networks, where a RSU is located at the place of interests, e.g., the campus entrance of a university. As illustrated in Fig. \ref{system_model}, the RSU by the side of the multi-lane road is actively pushing local informative contents such as high-resolution map and promotional video clip to the incoming vehicles.
Assuming that only the vehicles that enter the campus are interested in the distributed file, we focus on the performance of content distribution in the uni-direction.
Each vehicle carries a GPS device to calculate its geographical position, and each vehicle is equipped with two radios, which use control channel and service channel, respectively.
The exchange of the control packets helps the vehicles obtain the speed and position information of neighbors. So vehicles moving at relatively close speeds and within the same communication range can form a small cluster or a group. Vehicles with much higher or lower speeds will not be included in the same group\footnote{Such group assumption is valid in most of practical scenarios. Similar assumption and grouping technique have been applied in \cite{guo2017enabling,rossi2017stable}.}.
We assume a group has $k$ vehicles denoted by $V_1, V_2,\ldots,V_k$ from right to left. Their instantaneous moving speeds are $\upsilon$ in kilometers per hour with at most $\pm 5$km/h variations.
The average distance between any two vehicles $V_i$ and $V_{j}$  can be derived from the control packet and represented by $\bar{S}_{i,j}$, where $i, j\in\{1,2, \ldots, k\}$.
The communication range of the RSU is denoted by $R$. The reception probability beyond this range is assumed to be negligible. The length of the road that is within RSU's communication range is $L$.
\begin{table}
\caption{Related Notations.}
\label{Notation_table}
\centering
\begin{tabular}{|l||l|}
    \hline
    Notation & Definition \\ \hline
    $k$ & Vehicle group size \\ \hline
    $\gamma_{th}$ & Signal-to-noise ratio threshold \\ \hline
    $F$ & File size \\ \hline
    $\ell$ & Packet length \\ \hline
    $N$ & Total number of packets sent by the RSU \\ \hline
    $J$ & Total number of batches sent by the RSU \\ \hline
    $M$ & Batch size \\ \hline
    $R_b$ & Data transmission rate \\ \hline
    $P_t / P_t'$ & Transmission power at the RSU / vehicle \\ \hline
    $P_N$ & Noise power \\ \hline
    $f$ & Carrier frequency \\ \hline
    $m_1 / m_2$ & Fading shape factor for I2V / V2V channel \\ \hline
    $\beta$ & Path loss exponent \\ \hline
\end{tabular}
\vspace{-0.4cm}
\end{table}

As shown in Fig. \ref{system_model}, the content distribution consists of two phases. The first phase happens when the vehicles are passing through the coverage area of the RSU.
The reception of received packets depends on both the distance between the RSU and the receiving vehicle as well as the instantaneous channel gain. Since the distance between two adjacent vehicles is of much higher order than the wavelength, vehicles are assumed to have independent fading coefficients which follow the Nakagami-m distribution.
Due to the short connection period with the RSU and the impact of the lossy channel, a single vehicle may not be able to receive enough packets for decoding the original file.
Therefore, in addition to I2V communication, we introduce a broadcasting-based cooperative sharing method through V2V communication allowing vehicles to exchange the coded packets for further file recovery. These complementary transmissions in Phase 2 are initiated after the last vehicle leave the communication range of the RSU.
\subsection{Encoding and Decoding of BATS Code }
We assume the original file is divided into $F$ packets. In the first phase, the RSU uses the outer code of BATS code to generate coded packets in batch. To form a typical batch with index $j$, $\mathbf{d}_j$ packets are uniformly and randomly chosen from $F$ packets, and are linearly transformed into $M$ coded packets. This random degree $\mathbf{d}_j$ follows the pre-determined degree distribution $\Psi$. Thanks to the rateless nature of BATS code, the RSU can continuously and sequentially generate and broadcast BATS coded batches.
From the time the first vehicle $V_1$ enters the coverage area of the RSU until the last vehicle of the group leaves, the total number of packets sent by the RSU is denoted by $N$. As a batch consists of $M$ packets, equivalently, $J=N/M$ batches can be delivered to the vehicle group. We label each packet by packet index $n$, where $n=1,2,, \cdots N$, and the sequence number of the batches by batch index $j$, where $j=1,2,\cdots J$. So the $j$th batch contains the packets with indices from $(j-1)M+1$ to $jM$.
Denoted by $\ell$ the packet length in bits and $T_p\triangleq\ell/R_b$ the time for RSU sending one packet, where $R_b$ is the RSU transmission rate.

For cooperative transmissions in Phase 2, vehicles apply the inner BATS code which is the random linear network code to further encode the received packets of the same batch and broadcast.
With optimally designed BATS code, the original file can be decoded from any $F'=(1+\epsilon)F$ innovative packets\footnote{A packet is innovative if it is independent of all other packets.}, where $\epsilon\ll 1 $ is the coding overhead.
\section{Information Broadcast by RSU}\label{section_P1}
Based on our system model, the number of BATS coded batches that can be broadcast by the RSU is limited by the connection time between the RSU and the vehicles.
In this section, we analyze the theoretical number of innovative packets received by an individual vehicle and the whole group in Phase 1, respectively.
\subsection{Reception Status of a Single Vehicle}\label{section_single_vehicle}
We characterize the communication link between the RSU and the vehicle by a Nakagami-m fading channel. Since RSU is the only node that occupies the service channel for broadcasting data messages and all vehicles are in silent mode when receiving packets, there is no interfering signal posed by any vehicle. The signal outage probability of a vehicle is defined as the probability of its signal-to-noise ratio (SNR) dropping below the SNR threshold $\gamma_{th}$.
When the RSU starts to distribute the $n$th packet, the distance between the RSU and the vehicle $V_i$ is given by $d_{i,n}$.
We assume that the transmission time of a packet is sufficiently small so the distance variation during a single packet's transmission is negligible.
Given the RSU's transmission power $P_t$, noise power $P_N$ and path loss exponent $\beta$, the SNR $\gamma_{i,n}$ of vehicle $V_i$ when it starts receiving the packet $n$ can be expressed as
\begin{equation}
\gamma_{i,n} = \frac{P_t\cdot\alpha\cdot g\cdot \left(d_{i,n}\right)^{-\beta}}{P_N}
\end{equation}
where $g$ is the channel gain and the constant $\alpha$ denotes the power gain at the distance $d=1$m.
The channel gain $g$ of the Nakagami fading channel follows the gamma distribution with shape factor $m_1$ and average power $\Omega=\mathds{E}\left[g\right]$, i.e.,
\begin{equation}\label{f_g}
f_g(x)=(\frac{m_1}{\Omega})^{m_1}\frac{x^{m_1-1}}{\Gamma(m_1)}\exp\left(-\frac{m_1}{\Omega}x\right).
\end{equation}
We let $A(n,i)=\frac{\gamma_{th}}{\alpha}\frac{P_N}{P_t}\left(d_{i,n}\right)^{\beta}$. The signal outage probability of $V_i$ at the moment of starting receiving packet $n$ is given by
\begin{align}\label{equation_out_prob}
P_{i,n} &= \mathds{P}(\gamma_{i,n}<\gamma_{th}) \\ \nonumber
&=\mathds{P}(g<A(n,i)) \\ \nonumber
&=\int_{0}^{A(n,i)}f_g(x)dx \\ \nonumber
&=1-\frac{1}{\Gamma(m_1)}\Gamma\left(m_1,\frac{m_1}{\Omega}{A(n,i)}\right)
\end{align}
where $\Gamma(x)$ is the gamma function and $\Gamma(z,x)$ is the upper incomplete gamma function. Note that our study is not limited to the Nakagami-m fading channel. Any other channel model can be simply integrated into our proposed scheme by substituting the corresponding channel gain distribution $g$. Here we use the Nakagami fading channel since it has been widely applied in V2X research.

We use $P_i^{[j]}$ to represent the average signal outage probability of vehicle $V_i,i=1,2,\ldots,k$ when receiving the packets of batch $j$. We have
\begin{equation}\label{eqn:batch_loss_prob}
P_i^{[j]} =\frac{1}{M}\sum_{n=(j-1)M+1}^{jM}P_{i,n}
\end{equation} where $P_{i,n}$ is the outage probability for $V_i$ to start receiving packet $n$. The packets within each batch are sent consecutively by the RSU and hence the average SNR does not change much within the short transmission time of one batch. We assume the channel-induced packet loss probability is equivalent to the average signal outage probability. Therefore, the number of received packets of batch $j$ from the RSU can be approximated as a random variable following the Binomial distribution $\mathcal{B}(M,1-P_i^{[j]})$, with the average loss probability $P_i^{[j]}$. We denote the expectation value by $K_i^{[j]}$, then we have $K_i^{[j]}=M(1-P_i^{[j]})$. Hence the expected total number of packets received by a typical vehicle $V_i$ during the RSU broadcasting phase is given by $K_i=\sum_{j=1}^{J}K_i^{[j]}$.
\subsection{Reception Status of the Whole Group}
An innovative packet is received by the vehicle group when any one of the $k$ vehicles successfully receives it in Phase 1. Denote the expected number of innovative packets received by the vehicle group from batch $j$ by $K_g^{[j]}$. To find the expression for $K_g^{[j]}$, we first derive the packet loss probability for the vehicle group.

Consider a group of $k$ vehicles passing through the coverage area of a RSU. A packet is totally lost if none of the vehicles within group has successfully received it, which occurs with probability $\prod_{i=1}^{k}P_{i,n}$. Therefore, the probability that a packet from batch $j$ is not received by any vehicle is given by
\begin{equation}\label{eqn:group_loss_prob}
P_g^{[j]} = \frac{1}{M}\sum_{n=(j-1)M+1}^{jM}\left(\prod_{i=1}^{k}P_{i,n}\right)
\end{equation} where $P_{i,n}$ is the probability that vehicle $V_i$ fails to receive the $n$th packet. Accordingly, the expected number of innovative packets collected by the whole group from batch $j$ is $K_g^{[j]} = M\left(1-P_g^{[j]}\right)$. The total number of innovative packets received by the vehicle group during the RSU broadcasting phase is then given by $K_g = \sum_{j=1}^{J}K_g^{[j]}$.

When a vehicle is travelling at a relatively high speed, it may not receive sufficient packets for decoding the file during the short connection period with the RSU. On the other hand, the vehicle group has a longer connection period and also benefits from spatial diversity. Therefore, we usually have $K_g\gg K_i$, $\forall i$.
The margin between $K_g$ and $K_i$ provides an individual vehicle with the opportunity to receive more packets for decoding via the cooperative V2V transmissions, which are elaborated in the next section.
\section{Vehicle-to-Vehicle Cooperative Sharing}\label{section_P2}
After the vehicles leave the communication range of the RSU, the cooperative sharing is initiated to compensate the inadequate number of packets received from the RSU. In this section, we propose a utility-based method to obtain the scheduling at each individual vehicle in a distributed manner, which maximizes the effectiveness of each transmission and hence minimizes the total transmission delay.
\subsection{Distributed V2V Transmission}
To fully utilize the spatial and temporal diversity, the BATS inner code is applied for V2V communications. Specifically, when a vehicle has the chance to transmit, it will generate and broadcast a coded packet from all the received packets of a batch. The main design problem is the scheduling at each vehicle, i.e., which batch should be selected at a given time instance.
Generally, the proposed V2V-transmission algorithm consists of two steps: utility evaluation and distributed transmission scheduling. We elaborate on these two steps in the following sub-sections. In addition, the prior knowledge of each vehicle and the complete procedures of the proposed scheme are summarized in Algorithm 1.
\subsubsection{Utility evaluation}
When the last vehicle\footnote{A vehicle may identify itself as the last vehicle in a group if there is no other vehicle from the same group behind it. It can be recognized from the distance between two vehicles, which can be derived from the periodic broadcast control packet from peers.}
$V_k$ leaves the communication range of the RSU, it signals other vehicles to initiate the packet-sharing mechanism. Different from the existing literatures \cite{li2011codeon,yu2012rank}, our distributed method does not require the exchange of reception status, so the extra communication overhead is reduced. Therefore, after detecting the initialization signal, each vehicle evaluates the utility that can be brought to peer vehicles.

We use $\mathcal{N}_i^{[j]}$ to represent the set of packets of batch $j$ received by a typical vehicle $V_i$ from the RSU. The number of elements in set $\mathcal{N}_i^{[j]}$ is denoted by $Y_{i,j}$, i.e., $Y_{i,j}=|\mathcal{N}_i^{[j]}|$.
Let $y=|\mathcal{N}_i^{[j]}\backslash \mathcal{N}_{q}^{[j]}|$, where $y$ indicates the number of packets that are received by $V_i$ but not received by its neighboring vehicle $V_q$. Since each vehicle encodes all received packets of a batch and generates a certain number of coded packets in the V2V sharing phase, $y$ also represents the number of innovative coded packets of batch $j$ that $V_i$ can send to $V_{q}$. Though the packet set $\mathcal{N}_{q}^{[j]}$ is unknown to vehicle $V_i$, we can see that when $\mathcal{N}_i^{[j]}\nsubseteq \mathcal{N}_{q}^{[j]}$, $y\in(0,M]$. Based on the knowledge of $Y_{i,j}$ and the range of $y$, $V_i$ can estimate the amount of innovative information that is obtained by $V_q$ through each transmission of $V_i$. Here, we use utility to describe such amount of innovative information. As mentioned before, vehicle's critical information including position and speed is periodically broadcast via the control channel. The average distance between any two vehicles can be obtained, from which we can derive the packet loss probabilities between each pair of vehicles.
We assume that the V2V communication channel has the Nakagami fading with shape factor $m_2$. The transmission power of any vehicle is $P_t'$. Let $\alpha$ denote the power gain at the reference distance $d=1$m.
The SNR at receiver $V_{q}$ when $V_i$ is transmitting is given by
$\hat{\gamma_{i,q}} = \frac{P_t'}{P_N} g\alpha(\bar{S}_{i,q})^{-\beta}$, where $\bar{S}_{i,q}$ is the average distance between two vehicles.
Hence the packet loss probability from $V_i$ to $V_{q}$ is
\begin{align}\label{V2Vlossrate}
\hat{P_{i,q}}&=\mathds{P}(\hat{\gamma_{i,q}}<\gamma_{th}) \\\nonumber
& = 1-\frac{1}{\Gamma(m_2)}\Gamma\left(m_2,\frac{m_2}{\Omega}\frac{\gamma_{th}}{\alpha}\frac{P_N}{P_t'}(\bar{S}_{i,q})^\beta\right).
\end{align}
Meanwhile, in the process of receiving packets from the RSU in Phase 1, vehicle $V_i$ may build up the database of the channel environment.
We assume that each vehicle knows the packet loss probability from RSU to all its peers, $\{P_{i,n}, \forall {i, n}\}$, as well as that for the V2V channels, $\{\hat{P_{i,q}}, i,q = 1,...,k\}$. If $V_i$ has received $Y_{i,j}$ packets from the RSU, the probability that it can provide $V_q$ with $y$ innovative coded packets of batch $j$, which is also the probability that $y$ of $Y_{i,j}$ packets of batch $j$ were not successfully received by $V_q$ from the RSU, is given below
\begin{equation}\label{eqn:y}
\begin{aligned}
&\text{Pr}(y|Y_{i,j})\; \\
&=\begin{cases}
  \binom{Y_{i,j}}{y}\left(P_{q}^{[j]}\right)^y\left(1-P_{q}^{[j]}\right)^{Y_{i,j}-y}, &  y\leq Y_{i,j} \\
  0, & Y_{i,j}<y\leq M.
\end{cases}
\end{aligned}
\end{equation}

Since each batch contains at most $M$ innovative packets, we assume that the number of coded packets generated from each batch is no greater than $M$. Every time, network coefficients are randomly chosen from a finite field. The probability that different coded packets with the same batch index have dependent coefficients is negligible when the field size is sufficiently large, e.g., $GF(2^8)$. Hence we ignore the impact of network coefficients on the unsuccessful decoding throughout this paper.

Since $V_i$ can provide $V_{q}$ with only $y$ innovative packets for batch $j$, we know that beyond $y$ times of transmissions, the coded packet will no longer benefit $V_{q}$ unless some of the previous transmissions failed due to the channel-induced packet loss.
The more times of coded packets are broadcast, the less utility they can provide to the receiver.
In order to gain the maximal utility through a small number of V2V transmissions, it is important to know the transmission priority of a coded batch as well as to determine the number of transmissions of that batch. That is achieved by comparing utilities among different batches and different times of transmissions.

Now we calculate the probability that for a batch $j$, when $t$ V2V transmissions have been conducted, the next transmitted coded packet from $V_i$ is still innovative to $V_{q}$, which is denoted by $\text{Pr}(E_{i\rightarrow q}^{[j]}|t)$.
When the number of transmissions from $V_i$ to $V_{q}$ has not reached the number of innovative packets that $V_i$ possesses, i.e., $t<y$, the transmission will increase the total number of innovative packets at $V_{q}$ if it is successfully received. Otherwise, when $t\geq y$, the additional transmission may be useful only if $V_{q}$ has failed to receive all $y$ innovative packets from previous $t$ transmissions. Hence, we define two events:

$E_1=$\{The ($t+1$)th transmission is useful when at most $y-1$ coded packets were received in previous $t$ transmissions\}.

$E_2=$\{The ($t+1$)th transmission is useful regardless of the reception status of previous $t$ transmissions\}. \\
The probabilities of these two events are given below
\begin{equation}\label{Event}
\begin{aligned}
&\text{Pr}(E_1)=
& &\sum_{y=1}^{t}\text{Pr}(y|Y_{i,j})\sum_{l=0}^{y-1}\binom{t}{l}\left(1-\hat{P_{i,q}}\right)^l \\
&&&\times\left(\hat{P_{i,q}}\right)^{t-l} \\
&\text{Pr}(E_2)=& &\sum_{y=t+1}^{M}\text{Pr}(y|Y_{i,j}).
\end{aligned}
\end{equation}
The number of previous transmissions $t$ can be any value from $\{0,1,\ldots, M-1\}$. Thus, on condition that $V_i$ has transmitted $t$ times, the probability that the $(t+1)$th transmission of the coded packet of batch $j$, is still innovative to $V_q$ is expressed as
\begin{equation}
\text{Pr}(E_{i\rightarrow q}^{[j]}|t) = \text{Pr}(E_1)+\text{Pr}(E_2).
\end{equation}
This probability $\text{Pr}(E_{i\rightarrow q}^{[j]}|t)$ measures the amount of innovative information that $V_q$ will gain from the $(t+1)$th transmission of $V_i$, which is called utility in this paper. Hence, the total utility that all $k-1$ peers can obtain from this transmission is denoted by $\bar{U}(i,[j],t+1)$ and calculated by
\begin{equation}\label{Utility}
\bar{U}(i,[j],t+1)=\sum_{q\neq i}\text{Pr}(E_{i\rightarrow q}^{[j]}|t).
\end{equation}
The total utility $\bar{U}(i,[j],t+1)$ reflects the transmission priority of batch $j$.
However, some coded packets with different batch indices may have the same total utility if the above formula is in use. In order to differentiate the transmission sequence for those packets, we introduce some randomness by adding a random jitter $\varepsilon$. For not mixing up the random jitter with the utility differences, we amplify $\bar{U}(i,[j],t+1)$ by a value $\kappa$.
So we derive the final expression of the total utility
\begin{equation}
U(i,[j],t+1)=\kappa \bar{U}(i,[j],t+1)+\varepsilon.
\end{equation}
In practice, we choose $10^6$ for $\kappa$ and $\varepsilon$ is a uniformly distributed real number in the interval $(0,10)$, i.e., $\varepsilon=rand(0,10)$.
\subsubsection{Distributed transmission scheduling}
Vehicle $V_i$ locally calculates ${U}(i,[j],t+1)$ for different batches $j=1,2, \ldots, J$, and different times of transmissions $t+1=1,2, \ldots, M$. The coded packet of the batch that has larger utility should be given higher priority for transmission. Therefore, $V_i$ sorts these utilities in descending order. The corresponding sorted batch indices recorded in a row vector $\mathbf{R}_i$ indicate the scheduled transmission sequence of the coded packets. All $k$ vehicles follow the same rule to decide their transmission sequences of coded batches in a distributed manner.

To reduce the latency for transmitter selection, we adopt a random channel access protocol. This proposed protocol is also applicable for other channel access technology, such as the D2D communication in C-V2X. At the beginning of each slot, to avoid packet collisions, $k$ vehicles contend the channel with a random backoff delay which is randomly chosen from $[0, \Delta t_{max}]$, where $\Delta t_{max}$ is the maximum backoff delay. In order to accommodate more vehicles and ensure that the backoff difference can be recognized with the presence of signal propagation delay, we choose $\Delta t_{max}=50\mu s$.

When vehicle $V_i$ seizes the channel for broadcasting, it sequentially broadcasts one coded packet with the batch index as indicated in $\mathbf{R}_i$.
Whenever a coded packet is received, the vehicle will check whether it is independent of the previously received packets. If the coded packet cannot increase the rank of the decoding matrix of the corresponding batch, it will be dropped. A number of rounds of V2V transmissions are needed for all vehicles in the group to obtain $F$ innovative packets or achieve a total rank of $F$.
When a vehicle is able to decode the file, it piggybacks a stopping bit in its control packet to inform the peers. Once all vehicles have enough packets for decoding, the cooperative sharing stops.

\begin{algorithm}[t]
\small
{\caption{Distributed V2V transmission scheme}}
\begin{algorithmic}[1]
\STATE \textbf{Inputs:}\\
$J$: number of batches;\\
$M$: batch size;\\
$k$: number of vehicles;\\
$P_{i}^{[j]},\hat{P_{i,q}}, \forall i, q\neq i$: packet loss probability;\\
$\mathcal{N}_i^{[j]}$: set of received packets of batch $j$ in the I2V phase, only known by the corresponding vehicle.
\STATE \textbf{Outputs:} Batch transmission sequence $\mathbf{R}_i$ for $V_i, \forall i$.
\STATE \textbf{Step 1 utility evaluation}
\FOR{$i=1:k$}
\STATE$Y_{i,j}\gets|\mathcal{N}_i^{[j]}|$ \\
       \FOR{times of transmissions $t=0:M-1$}
       \FOR{$q=1:k,\  q\neq i$}
        \STATE$y$ denotes the number of innovative packets sent by $V_i$ to $V_q$;
       \FOR{$y=1:t$}
       \STATE$\text{Pr}(y|Y_{i,j})$ is calculated according to (\ref{eqn:y});\\
       \STATE$\text{Pr}(E_1)$ is calculated according to (\ref{Event});
       \ENDFOR
       \FOR{$y=t+1:M$}
       \STATE$\text{Pr}(y|Y_{i,j})$ is calculated according to (\ref{eqn:y});\\
       \STATE$\text{Pr}(E_2)$ is calculated according to (\ref{Event});
       \ENDFOR \\
       \STATE$\text{Pr}(E_{i\rightarrow q}^{[j]}|t)\gets\text{Pr}(E_1)+\text{Pr}(E_2)$
       \ENDFOR \\
       \STATE The total utility of $V_i$ when transmitting the coded packet of batch $j$ for the $(t+1)$th transmission is: \\$U(i,[j],t+1)\gets\kappa\sum_{q\neq i}\text{Pr}(E_{i\rightarrow q}^{[j]}|t)+\varepsilon$
       \ENDFOR
\ENDFOR
\\
\STATE \textbf{Step 2 distributed transmission scheduling}
\FOR{i=1:k}
\STATE $V_i$ sorts $U(i,[j],t+1)$ for $j=1, 2, \ldots, M$ and $t+1=1, 2, \ldots, M$ in descending order; \\
\STATE $V_i$ creates a row vector $\mathbf{R}_i$ containing the corresponding sorted batch indices.
\ENDFOR
\end{algorithmic}
\end{algorithm}


\subsection{Lower Bound of the Total Number of Transmissions}\label{section_lower_bound}
The lower bound of the total number of V2V cooperative transmissions is obtained when there is an ideal packet reception condition in Phase 1 such that every packet received by a vehicle from the RSU can contribute to all other neighboring vehicles for the file decoding. So in the V2V cooperative sharing phase, assuming a vehicle $V_q$ will broadcast $X_q$ coded packets, then each coded packet is innovative to all previously received packets by any other vehicle $V_i, i\neq q$. That means $\bar{U}(q,[j],t+1)$ in (\ref{Utility}) is equal to $k-1$ for all possible $t$ values. From a receiver $V_i$'s perspective, based on the number of packets collected from the RSU in Phase 1, we know that at least $F-\sum_{j=1}^{J}Y_{i,j}$ innovative packets should be received from the rest vehicles. Given the average packet loss probability calculated by (\ref{V2Vlossrate}), we can form an inequality relationship between $X_q$ and $F-\sum_{j=1}^{J}Y_{i,j}$. To obtain the lowest number of V2V transmissions, we need to find the minimum value of the summation of transmissions at each vehicle, $X_i, i =1,2,\ldots,k$. It is equivalent to solving the following linear optimization problem.
\begin{equation}\label{optimization}
\begin{aligned}
& \text{minimize}
& & \sum_{i=1}^{k}X_i \\
& \text{subject to}
& & \sum_{q\neq i }\left(1-\hat{P_{q,i}}\right)X_{q} \geq F-\sum_{j=1}^{J}Y_{i,j}, \; \\
&&& 0\leq X_i \leq\sum_{j=1}^{J}Y_{i,j} , \; i = 1,2, \ldots,k.
\end{aligned}
\end{equation}
The solution of above problem defines the lower bound of the total number of V2V cooperative transmissions. Given packet length $\ell$ and maximum backoff delay $\Delta t_{max}$, the lower bound of the V2V cooperative sharing delay can be obtained.

However, the lower bound can hardly be achieved in reality because packets received during the RSU broadcasting phase are totally random and can be highly correlated as a result of short inter-vehicle distance. So the transmitted coded packets cannot always be innovative. Extra cooperative transmissions are required to ensure every vehicle successfully decodes the file. Nevertheless, this lower bound will provide a good reference for us to evaluate the effectiveness of the proposed scheduling method.
\section{Determination of Rank Distribution for BATS Code Design}\label{section_rank}
The design of degree distribution $\Psi$ of BATS code at the source node is critical to achieving high coding efficiency at the end of content distribution \cite{xu2017quasi}.
The optimal degree distribution can be obtained through a linear optimization problem defined in \cite{yang2014batched,ng2013finite}, given the channel rank distribution, which is defined as the probability of the rank for each batch. Since our network involves cyclic topology, and the transmission priority is assigned according to the utility, resulting in a complicated analysis of the rank distribution. In this section, some approximations are adopted to obtain an accurate estimation of the channel rank distribution in a simplified way.

The last vehicle to successfully decode the file is referred to as the bottleneck vehicle.
Since other vehicles are guaranteed to decode the file if the bottleneck vehicles can decode, we optimize the BATS code to ensure the bottleneck user can decode the file with the minimum number of V2V transmission.
So in this section, we focus on the rank evaluation of the bottleneck vehicle at the moment of successfully decoding the original file.

In the broadcasting phase, all the vehicles have roughly the same connection time with the RSU and hence they will receive the similar number of packets from the RSU. In the cooperative sharing phase, based on the random channel access protocol, each vehicle has an approximately equal transmission opportunity. However, due to the poorer channel connection with peers caused by longer average transmission distance, the front and back vehicles in the group experience larger packet loss probability when receiving coded packets from peers. Therefore, any of them is more likely to be the bottleneck user of the network, as shown in Fig. \ref{system_model}.
Nevertheless, without loss of generality, we consider the bottleneck vehicle could be any one of $k$ vehicles and denote this bottleneck vehicle by $V_b$.
Based on the expected number of received packets from the RSU, we know that the bottleneck vehicle requires $\Delta=F-K_b$ innovative packets to be contributed by the rest of $k-1$ vehicles. To estimate the rank distribution, we need to know how these $\Delta$ packets are distributed over batches and $(k-1)$ vehicles according to the proposed utility-based transmission method.
So our analysis comprises two parts: the probability of sending a particular batch which is determined by the utility, and the amount of innovative content of each batch generated by a vehicle.
\subsubsection{Batch selection probability}
For a typical vehicle, a batch with larger total utility for a single transmission is assigned with higher broadcasting priority. It indicates that the probability of sending a certain batch is proportional to the total number of innovative coded packets provided by that batch to peers.
Coded packets generated by a non-bottleneck vehicle $V_i$ $(i\neq b)$ with batch index $j$ are innovative to a peer vehicle $V_{q}$ if they are formed by the packets that are only received by $V_i$ from the RSU.
We use set $\mathcal{I}_{i,j}(q)$ to record these innovative packets. Hence based on the understanding of (\ref{eqn:batch_loss_prob}) and (\ref{eqn:group_loss_prob}), the expected number of elements in set $\mathcal{I}_{i,j}(q)$ can be derived as below
\begin{equation}
\left|\mathcal{I}_{i,j}(q)\right|=\sum_{n=(j-1)M+1}^{jM}\left(1-P_{i,n}\right)P_{q,n}.
\end{equation}
Denote by $\rho(i,j)$ the probability that for a single transmission, $V_i$ selects the coded packet with batch index $j$ for V2V transmission. Hence $\rho(i,j)$ is defined by the ratio between the total utility of that particular batch and all batches, i.e.,
\begin{equation}\label{batch_selection_prob}
\rho(i,j)=\frac{\sum_{q\neq i}\left|\mathcal{I}_{i,j}(q)\right|}{\sum_{j=1}^{J}\sum_{q\neq i}\left|\mathcal{I}_{i,j}(q)\right|}.
\end{equation}
\subsubsection{Amount of innovative content}
In this part, we are going to analyze the total amount of innovative content that the peer vehicles can offer the bottleneck vehicle. In general, $V_b$ can be any vehicle from $V_1$ to $V_k$. For convenience of presentation, we use a simple mapping to represent the identifiers of the vehicles excluding the bottleneck vehicle by adopting the function $w(s),s=1,2,\ldots,k-1$. The mapping rule is as follows:
\begin{equation}
w(s)=
\begin{cases}
s, & s<b \\
s+1, & s\geq b.
\end{cases}
\end{equation}
For instance, if $b=2$, the vehicles except the bottleneck vehicle $V_b$ are $V_1,V_3,\ldots,V_k$. They can be represented by $V_{w(1)},V_{w(2)},\ldots,V_{w(k-1)}$ correspondingly.

From the previous part, we know that a non-bottleneck vehicle $V_{w(s)}$ can provide $V_b$ with $\left|\mathcal{I}_{w(s),j}(b)\right|$ innovative packets for batch $j$.
In order to find the total number of innovative packets that can be provided from $k-1$ peer vehicles to $V_b$ for a typical batch $j$, we need to find the number of elements in the union of $k-1$ sets, i.e., $\left|\cup_{s=1}^{k-1} \mathcal{I}_{w(s),j}(b)\right|$. This requires the elimination of the duplicate elements which appear in more than one set. Since the common elements that can appear in two sets, three sets, \ldots, and at most $k-1$ sets, we introduce a general formula to calculate the number of common elements in $m$ sets. Given that $2\leq m\leq k-1$ and $1\leq u_1<u_2<\ldots <u_m\leq k-1$, we have
\begin{equation}
\begin{aligned}
&\left|\mathcal{I}_{w(u_1),j}(b)\cap\ldots\cap \mathcal{I}_{w(u_m),j}(b)\right| \\
&=\sum_{n=(j-1)M+1}^{jM}\left[\prod_{l=u_1}^{u_m}\left(1-P_{w(l),n}\right)P_{b,n}\right].
\end{aligned}
\end{equation}

In each time slot of V2V transmissions, all $k$ vehicles contend the channel with randomly selected backoff times. Therefore, every vehicle has an approximately equal opportunity to access the channel. For each transmission opportunity, the batch selection probability has been given in (\ref{batch_selection_prob}). Thus, we can calculate the amount of innovative content for a typical batch $j$ that is delivered to the bottleneck vehicle when the V2V transmission is conducted once, denoted by $I(j)$. According to the Inclusion-Exclusion Principle as well as above analyses, the expression of $I(j)$ is shown in (\ref{I_equation}).
\begin{figure*}
\begin{equation}\label{I_equation}
\begin{aligned}
I(j)
&=\sum_{s=1}^{k-1}\rho(w(s),j)\cdot\left|\mathcal{I}_{w(s),j}(b)\right|+\sum_{m=2}^{k-1}(-1)^{m+1} \\
&\times\left\{\sum_{1\leq u_1<\ldots <u_m\leq k-1}\min\left[\rho\left(w(u_1),j\right),\ldots,\rho\left(w(u_m),j\right)\right]\cdot\left|\mathcal{I}_{w(u_1),j}(b)\cap\ldots\cap \mathcal{I}_{w(u_m),j}(b)\right|\right\}.
\end{aligned}
\end{equation}
\hrulefill
\end{figure*}
Hence we determine a cutoff integer $c$ to have
\begin{equation}
\Delta \approx c\cdot\sum_{j=1}^{J}{I(j)}.
\end{equation}
Therefore, the number of innovative packets of each batch that are contributed by $k-1$ peers to the bottleneck vehicle is $cI(j),\forall j$. Since $V_b$ in theory receives $K_b^{[j]}$ packets from the RSU, the expected number of innovative packets or the rank of batch $j$ is given by $K_b^{[j]}+cI(j)$.
Let $r$ be the average rank of a batch, where $r=0,1,\ldots M$. Knowing the expected rank value of each batch, the equivalent probability that the batch $j$ has an innovative packet to increment its rank by the end of the content distribution is $p_{e}^{[j]}=\left[K_b^{[j]}+cI(j)\right]/M$. As a result, the estimated cumulative distribution function (CDF) of rank $r, r\in\{0,1,\ldots,M\}$, is calculated by
\begin{equation}
F_e(r)=\frac{1}{J}\sum_{v=0}^{r}\sum_{j=1}^{J}\binom{M}{v}\left(p_{e}^{[j]}\right)^v\left(1-p_{e}^{[j]}\right)^{M-v}.
\end{equation}
By converting $F_e(r)$ into the probability density function, we can obtain the estimated rank distribution $f_e(r)$ to ensure the successful decoding at the bottleneck vehicle. Note that with the same degree distribution, other vehicles are guaranteed to decode the file successfully since they receive more innovative packets.
\section{Performance Evaluation}\label{section_simulation}
In this section, numerical results are given to validate the analytical results and evaluate the performance of the proposed V2X content distribution scheme.
\subsection{Simulation Settings}
Our simulation environment is set up according to the content distribution system model described in Section \ref{system} where a RSU is located at the entrance of a campus, distributing digital maps to the incoming vehicles. For simplicity, we consider a straight road consisting of two lanes with lane width of 3m. The RSU is $50$m away from the centre of the nearest lane. The heights of the RSU and each vehicle are 8m and 1m. The communication range of the RSU as well as the vehicle is 200m. The large-scale path loss is characterized by the dual-slop model \cite{cheng2007mobile}:
\begin{equation}
PL=
\begin{cases}
PL_{0}+10\beta_1 log_{10}{\frac{d}{d_0}}, & d_0<d\leq d_c \\
PL_{0}+10\beta_1 log_{10}{\frac{d_c}{d_0}}+10\beta_2 log_{10}{\frac{d}{d_c}}, & d>d_c
\end{cases}
\end{equation}
where $PL_0$ is the free space path loss at the reference distance $d_0=10$m, and the critical distance $d_c=80$m.
We assume the line-of-sight (LOS) is the dominant component in Phase 1. So we choose Nakagami fading factor $m_1=1.2$ for characterizing the I2V channel. On the other hand, in V2V communication, vehicles encounter more signal reflections for the vehicles separated by longer distance. We choose $m_2$ to be $1.2$ when the distance between any two vehicles is less than $90$m whereas $0.75$ for the distance beyond $90$m. Other parameter values are shown in Table \ref{Value_table}. In this paper, we directly use MATLAB tool to generate the network model and the traffic model because we mainly focus on the content downloading and sharing among a group of vehicles. Our proposed scheme has considered the variations of the inter-vehicle distances and speeds.
\begin{table}
\caption{Parameter Settings in Simulations.}
\label{Value_table}
\centering
\begin{tabular}{|l|l|} \hline
    File size $F$ & $12000$ packets \\ \hline
    Packet length $\ell$ & $1500$ Bytes \\ \hline
    Data transmission rate $R_b$ & $6$ Mbps \\ \hline
    Batch size $M$ & 16 \\ \hline
    Maximum random backoff delay & $50\mu$s \\ \hline
    Transmission power $P_t, P_t'$ & $20$ dBm, $20$ dBm \\ \hline
    Noise power $P_N$ & -89 dBm \\ \hline
    SNR threshold $\gamma_{th}$ & 10 dB \\ \hline
    Carrier frequency $f$ & $5.9$ GHz \\ \hline
    Path loss model & \specialcell{Dual-slop model\cite{cheng2007mobile}:\\ $\beta_1=2.3; \beta_2=2.7$} \\ \hline
    Fading model& \specialcell{Nakagami fading:\\ $m_1=1.2$; $m_2=1.2,0.75$} \\ \hline
    Height of RSU, vehicle & 8m, 1m \\ \hline
    Lane width & 3m \\ \hline
    \end{tabular}\vspace{-0.0cm}
\end{table}
\subsection{Simulation Results and Discussions}
\subsubsection{Phase 1 packet reception status and packet gain}
\begin{figure}
\centering
\includegraphics[width=0.47\textwidth]{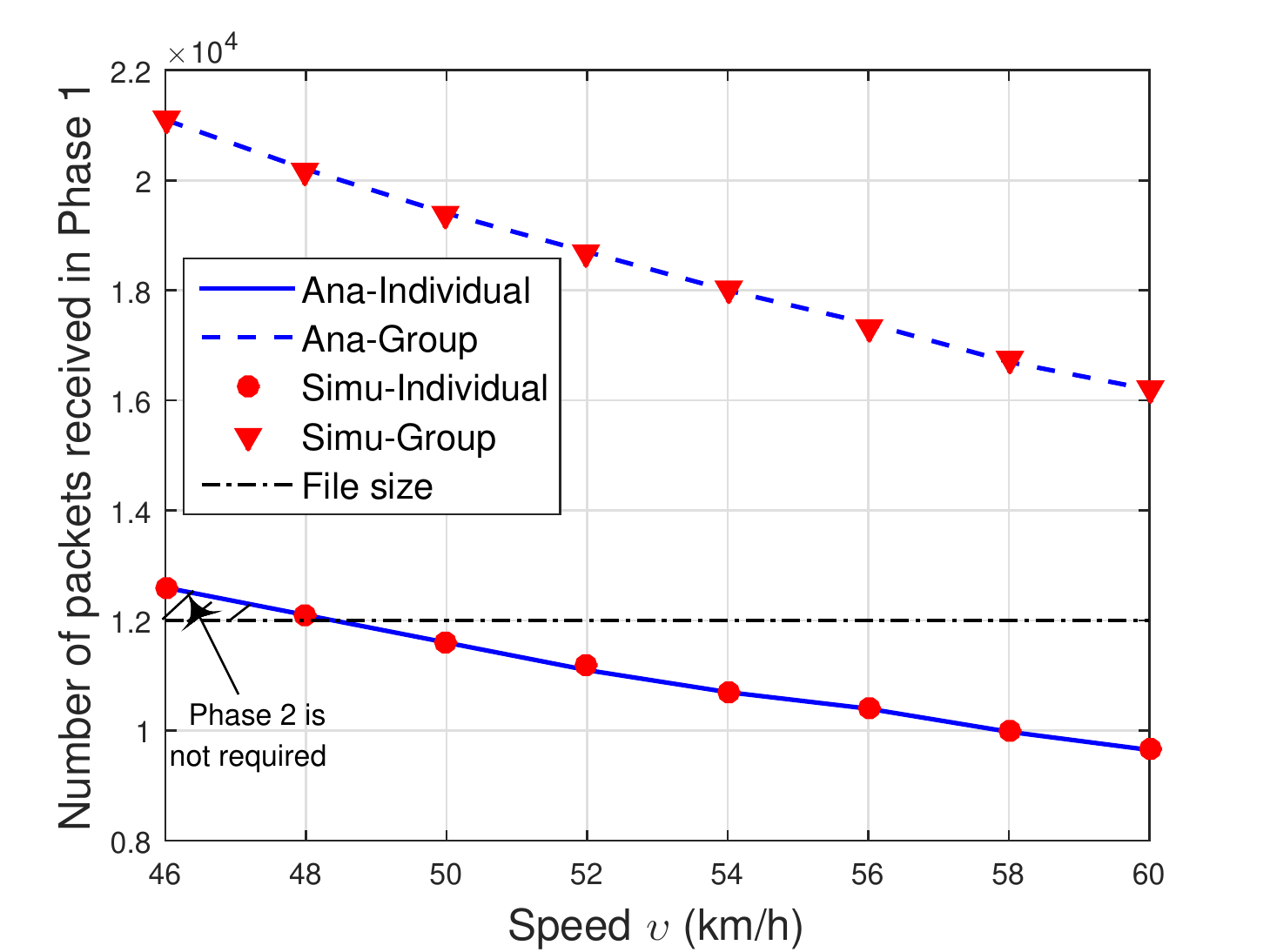}
\vspace{-0.2cm}
\caption{The impact of the average speed $\upsilon$ on Phase 1 packet reception status.}
\label{speedgain}
\end{figure}

\begin{figure}
\centering
\includegraphics[width=0.47\textwidth]{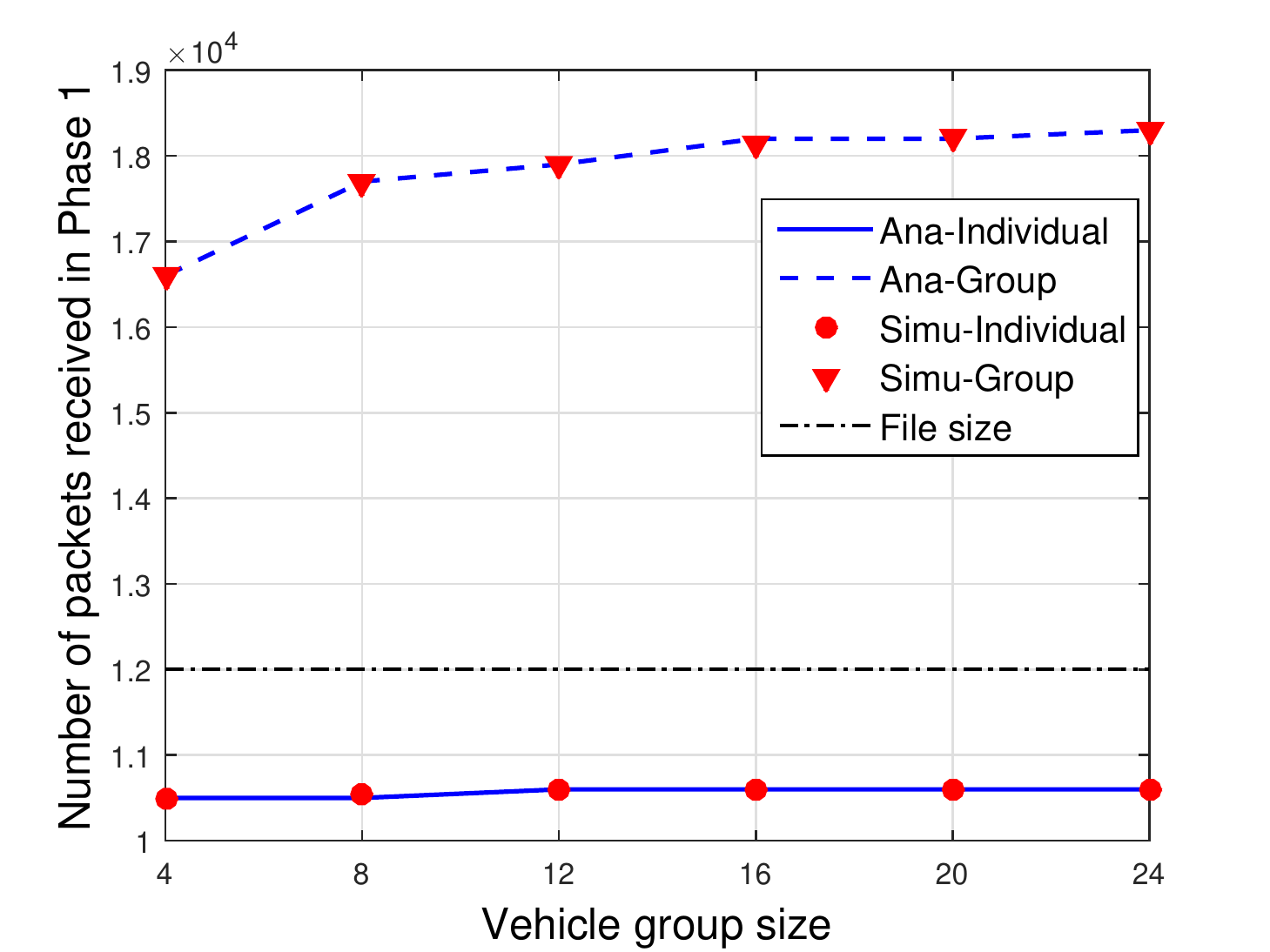}
\vspace{-0.2cm}
\caption{The impact of the vehicle group size on Phase 1 packet reception status.}
\label{groupsizegain}
\end{figure}

We assume that the vehicles with relatively close moving speeds and within the same communication range form a group. Similar to \cite{wang2012joint}, at any time instance, a typical vehicle $V_i$ moves at $\upsilon$ in kilometers per hour, with at most $\pm 5$km/h variation. The horizontal separation between any two adjacent vehicles is randomly distributed. Fig. \ref{speedgain} and Fig. \ref{groupsizegain} verify the theoretical analysis of the packet reception status of Phase 1. In Fig. \ref{speedgain}, eight vehicles form a group moving at different average speeds $\upsilon$ and a variation up to $\pm 5$km/h. Based on the current simulation settings, when $\upsilon\leq48$ km/h, the cooperative sharing is not needed since each vehicle can receive sufficient number of packets in phase 1. Fig. \ref{groupsizegain} is obtained when the vehicles move at $55$km/h$\pm 5$km/h. With more number of vehicles in the group, an individual vehicle can achieve a higher packet gain from the group.
\subsubsection{Accuracy of the estimated rank distribution}
\begin{figure}
\centering
\includegraphics[width=0.47\textwidth]{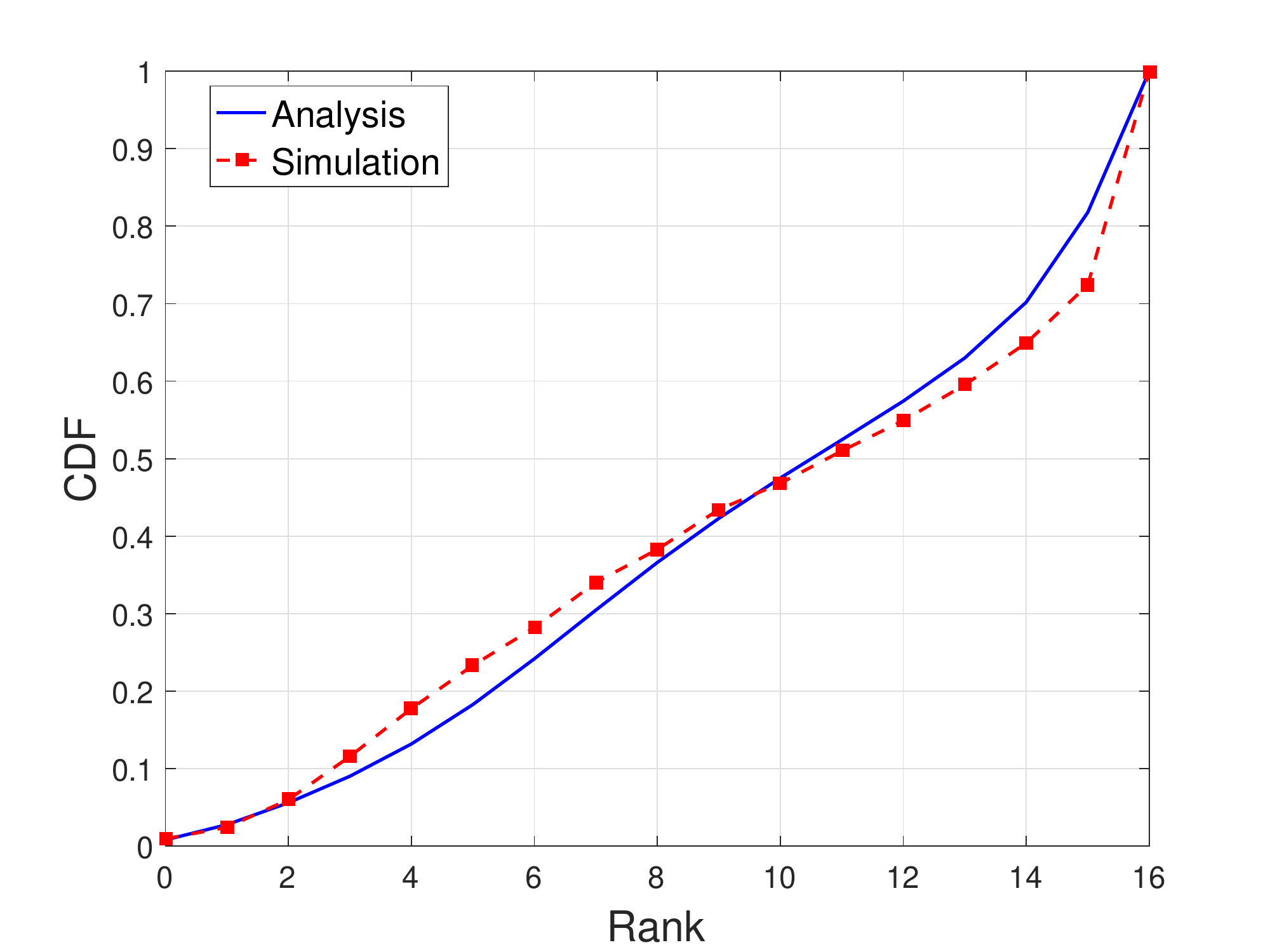}
\vspace{-0.2cm}
\caption{Rank CDF. Vehicle group size: $4$. Speed: $55$km/h$\pm 5$km/h.}
\label{cdfspeed55}
\end{figure}

\begin{figure}
\centering
\includegraphics[width=0.47\textwidth]{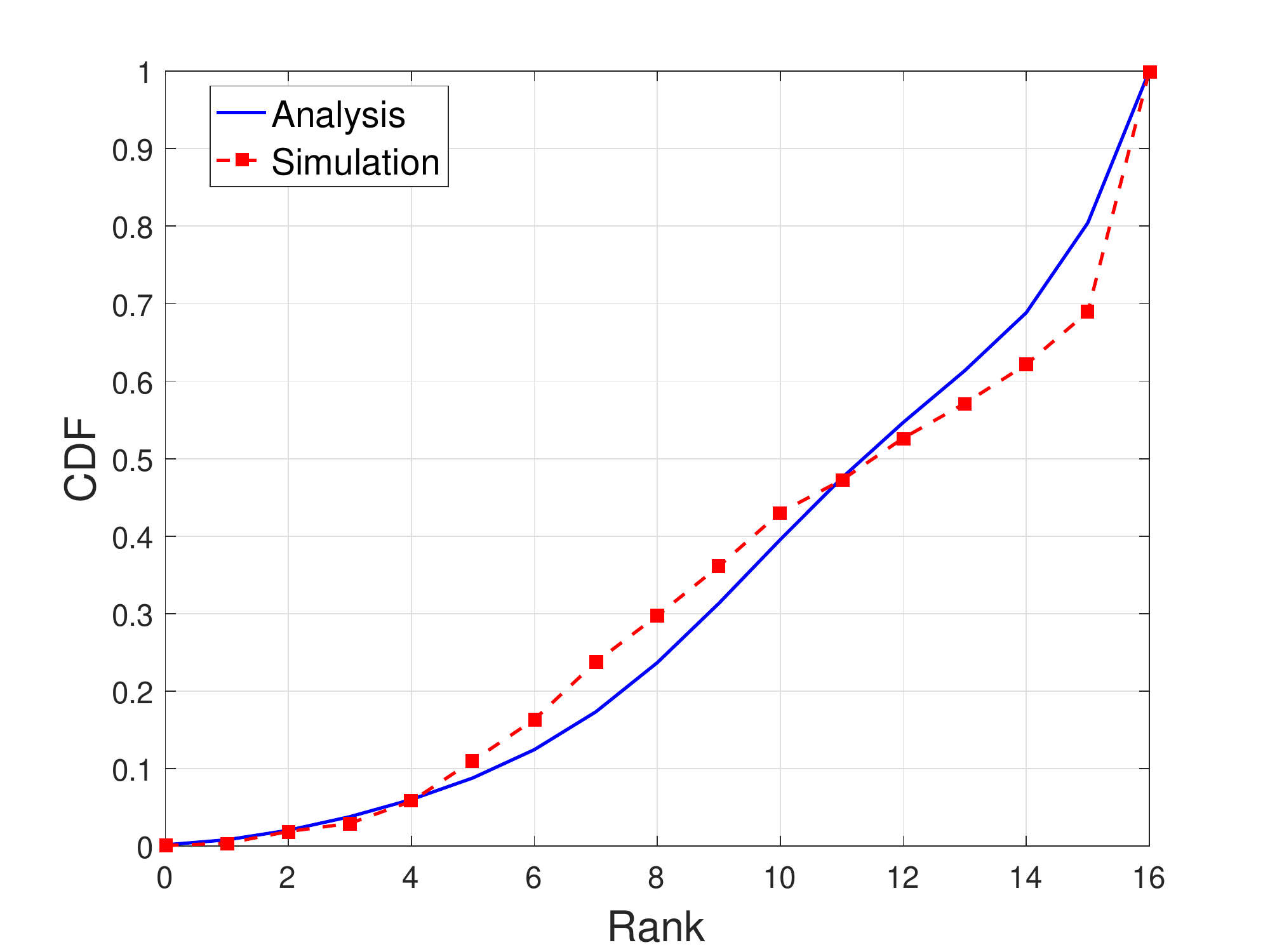}
\vspace{-0.2cm}
\caption{Rank CDF. Vehicle group size: $4$. Speed: $60$km/h$\pm 5$km/h.}
\label{cdfspeed60}
\end{figure}

The effectiveness of the estimated rank distributions which are derived in Section \ref{section_rank} is proved in Fig. \ref{cdfspeed55} and Fig. \ref{cdfspeed60} given two different average vehicle speeds. From the numerical results, vehicle $V_1$ is the bottleneck vehicle, hence the performance of $V_1$ is observed. The analytical rank CDF plotted in blue match the simulation results closely. Intuitively, we can learn that by adopting BATS code, it is not necessary for the vehicle to achieve the full rank of each batch so that the original file can be decoded.
The slop of the curve in Fig. \ref{cdfspeed60} is deeper than that in Fig. \ref{cdfspeed55} because the packet gain from the group is less so more batches have to reach full rank for successful decoding.
By converting the analytical CDF to the rank distribution, it is observed that the analytical distributions introduce only $5.3\%$ and $5.4\%$ coding overhead for the degree distribution optimization, which confirms the effectiveness of code design based on the analytical rank distribution.
\subsubsection{Throughput of the proposed scheme}
\begin{figure}
\centering
\includegraphics[width=0.47\textwidth]{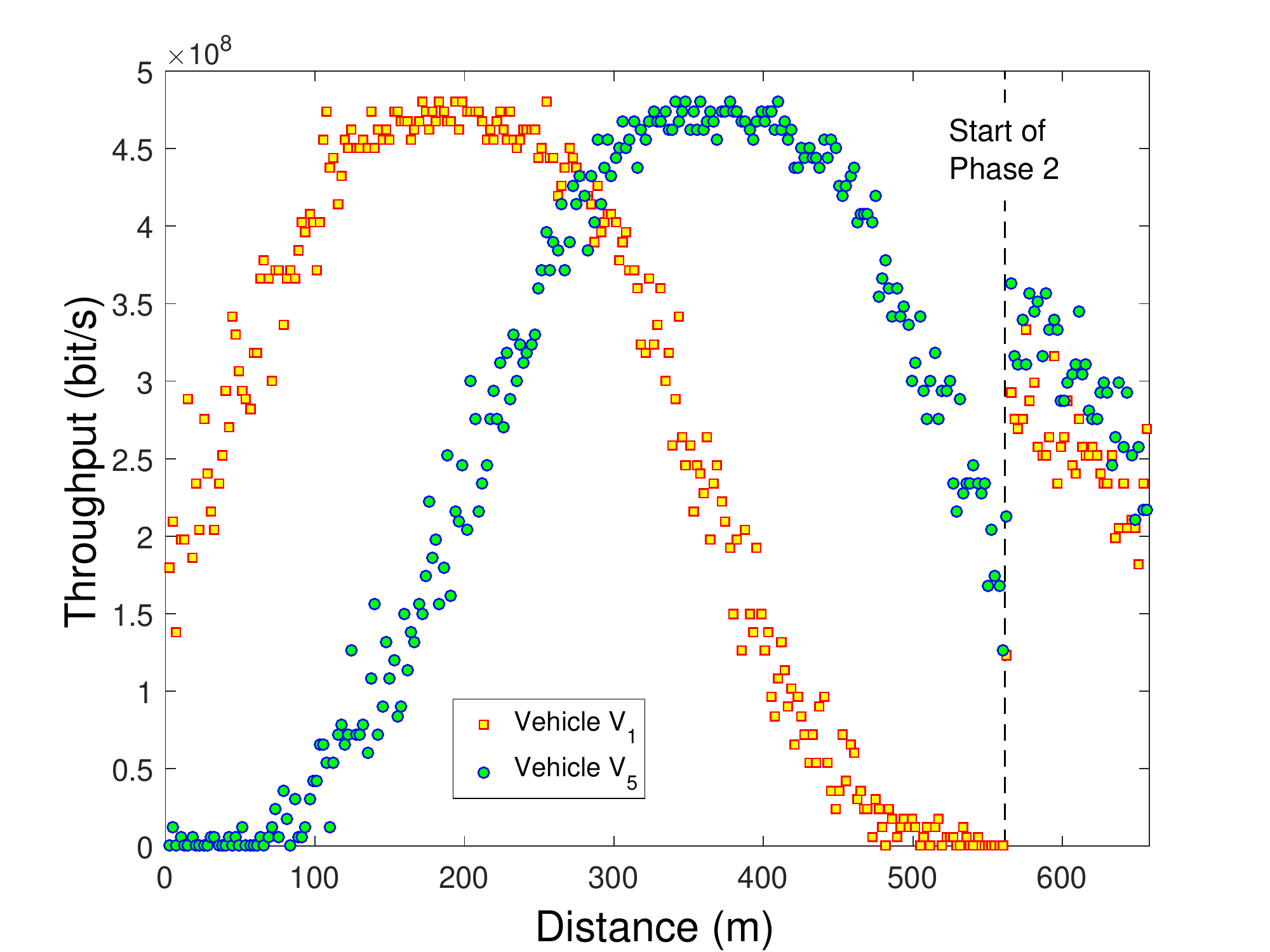}
\vspace{-0.2cm}
\caption{Throughput of vehicles $V_1$ and $V_5$ versus $V_1$'s moving distances. Vehicle group size: $8$. Speed: $55$km/h$\pm5$km/h.}
\label{Throughput}
\end{figure}

We consider a group of eight vehicles, namely $V_1$ to $V_8$, according to the sequence of passing through the communication range of the RSU.
Based on the numerical results, we find that vehicle $V_1$ is the bottleneck vehicle while vehicle $V_5$ is the first one to start decoding. The throughput performances of these two vehicles are shown in Fig. \ref{Throughput}. Regarding the point that $V_1$ just enters the communication range of the RSU as the origin, the horizontal distances between $V_1$ and the origin are represented by the X-axis values. When the vehicle moves closer to the RSU, the throughput increases due to the shorter communication distance but decreases after passing by the RSU due to the larger communication distance. $V_1$ and $V_5$ are separated by a certain distance so an obvious throughput shift is observed in the figure. Note that when vehicles are conducting the second-phase communication by following the proposed V2V sharing scheme, the throughput will drastically increase, which validates the effectiveness of our proposed scheme.
\subsubsection{Performance benchmarking}
We compare our content distribution scheme with two data dissemination frameworks in VANETs. One is a popular scheme called CodeOnBasic \cite{li2011codeon} which is the packet-level version of CodeOn, and the other one is a network coded version of RLS-SNR \cite{zeng2017channel} which is a centralized scheme that uses RLS for channel prediction and SNR-based utility for V2V scheduling.
Generally, our scheme outperforms two benchmark schemes in three major aspects:
\begin{itemize}
\item Our proposed scheme exploits the rateless property of BATS code so the vehicle group is not required to collect all packets of each batch from the RSU. In contrast, the RSU in CodeOnBasic as well as NC-based RLS-SNR divides the file into $F/M$ blocks and performs RLNC within each block of $M$ packets. The vehicle group is required to collect all the coded packets of each block for the next-phase V2V sharing.
\item When vehicles work cooperatively to share the received content, without exchanging the reception status, the proposed scheme has a good total utility evaluation method that can be conducted distributedly at each vehicle. CodeOnBasic relies on the frequent update of the rank of each block and selects a new transmitter based on the backoff delay up to $2ms$. NC-based RLS-SNR requires the update of decoding status of each vehicles to the central control after a round of transmissions and the scheduling decision is sent back from the control to vehicles. Both schemes introduce extra network overhead.
\item To determine the transmission sequence, CodeOnBasic uses the rank difference between two vehicles as amount of innovative information. It may easily cause the wrong termination of the sharing phase when the number of coded packets of any block for two vehicles are the same but smaller than the block size. NC-based RLS-SNR ignores the impact of the fading channel on the packet reception at each vehicle thus the computation of the packet utility may not accurately reflect the decoding status. Therefore, these two schemes for transmission scheduling may end up with sub-optimal solutions. Our proposed scheme perfectly avoids above problems by using BATS code and taking both packet loss probability and utility of the coded packets into account.
\end{itemize}

\begin{figure}
\centering
\includegraphics[width=0.47\textwidth]{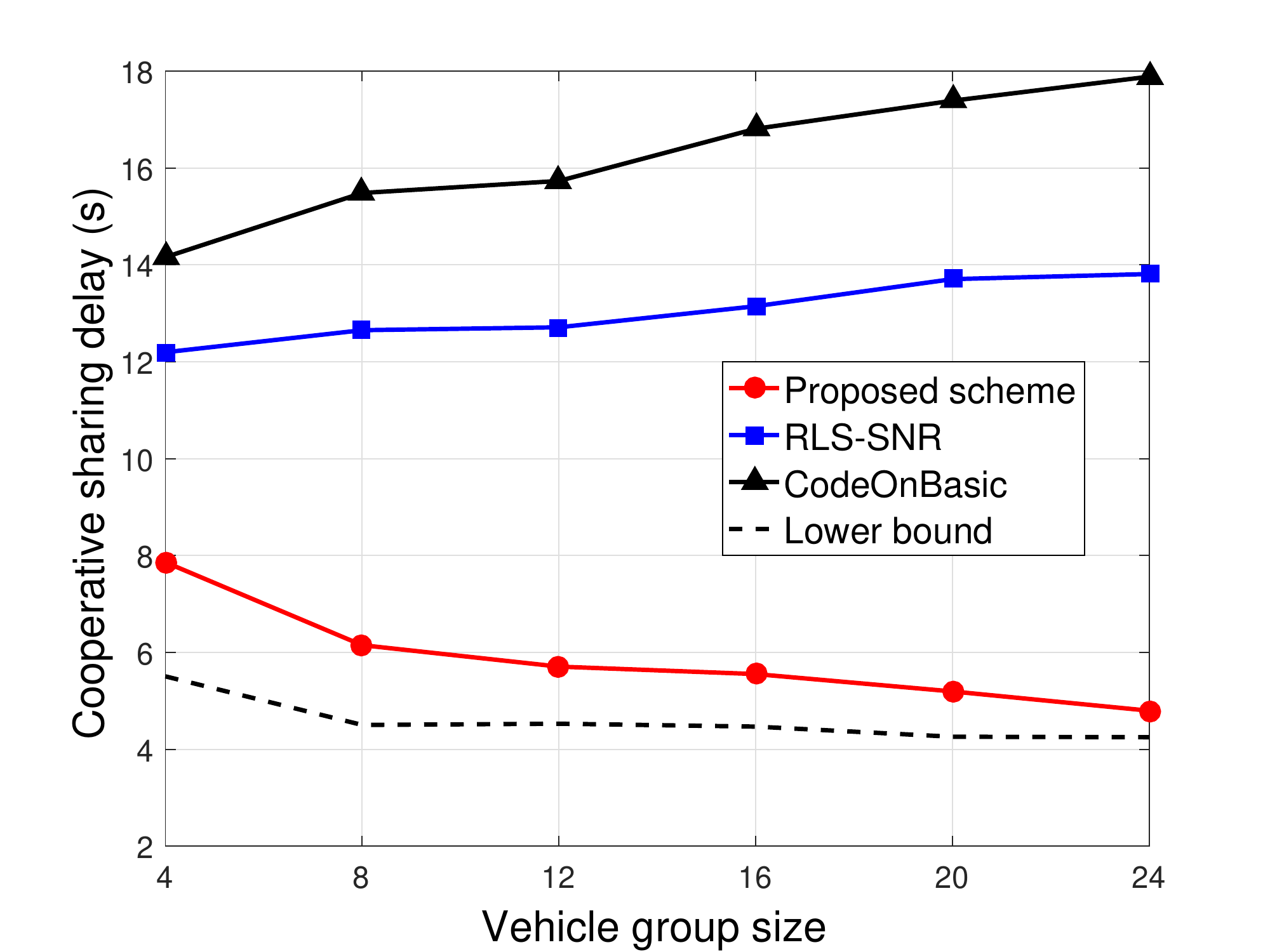}
\vspace{-0.2cm}
\caption{The impact of vehicle group size on Phase 2 delay.}
\label{groupsize}
\end{figure}

The performances of the V2V sharing delay by using our proposed scheme, NC-based RLS-SNR, and CodeOnBasic are evaluated in Fig. \ref{groupsize} where the vehicle speed is $55$km/h$ \pm 5$km/h. When the group size expands from 4 to 24, the V2V transmission delay of the proposed scheme is observed to decline and get closer to the lower bound because more innovative packets that the vehicle group can obtain from the RSU. On the other hand, the benchmark schemes have an obvious delay increase because they are constrained by the requirement of receiving $M$ coded packets per block for file decoding.
The duration of Phase 1 is 36.7s, same for different schemes. To depict the overall distribution delay, all curves will be shifted upward with 36.7 units.

\begin{figure}
\centering
\includegraphics[width=0.47\textwidth]{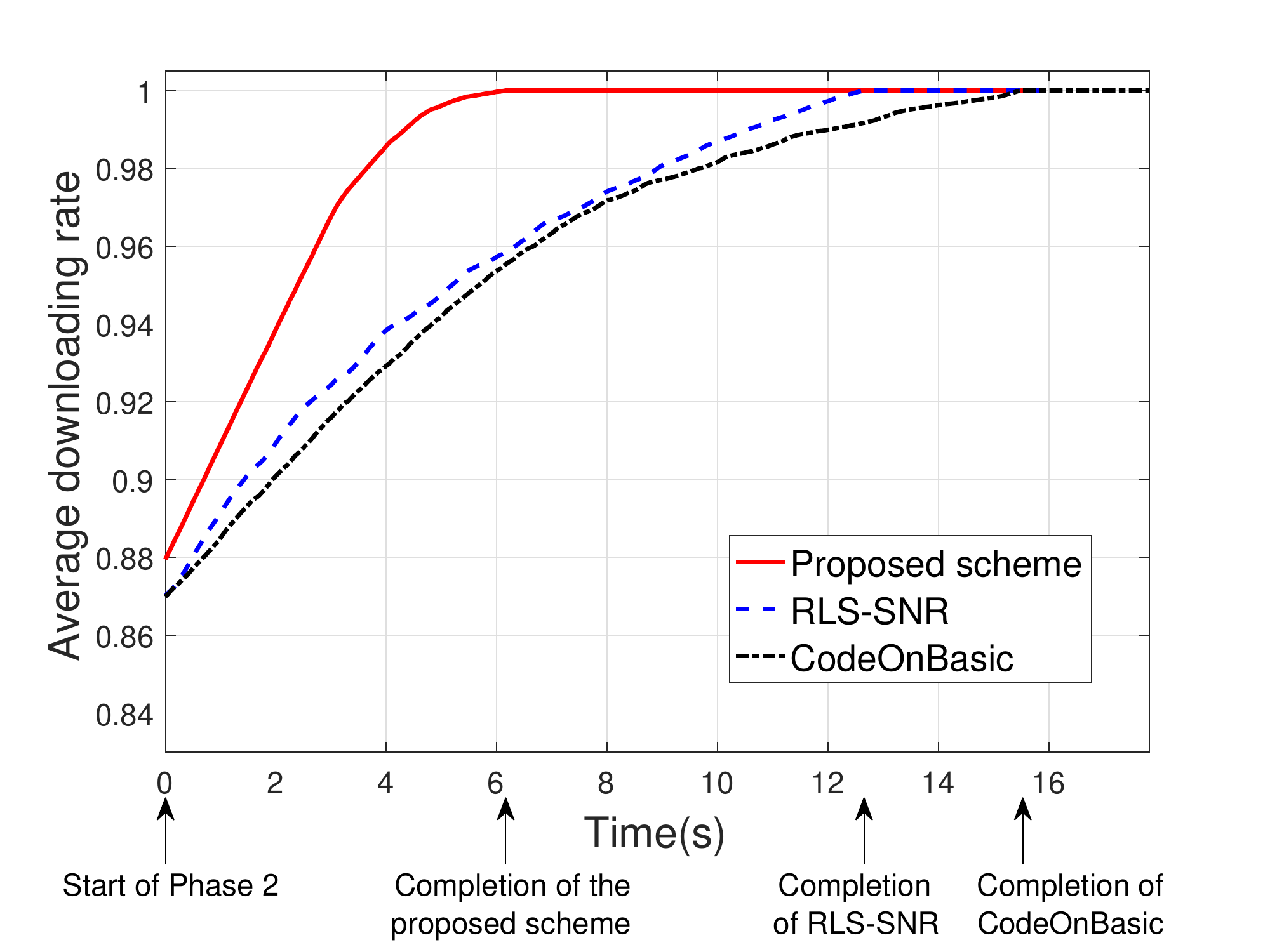}
\vspace{-0.2cm}
\caption{Average downloading rate.}
\label{downloading}
\end{figure}

Fig. \ref{downloading} shows the comparison between the proposed scheme and two benchmark schemes regarding the average downloading delay of a single vehicle, given vehicle group size of 8 and speed at $55$km/h$\pm 5$km/h. Note that under the same simulation setting, CodeOnBasic and NC-based RLS-SNR cannot guarantee that the vehicle group can receive $M$ coded packets for each block from the RSU due to the fading channel. So the RSU is adjusted to send one redundant coded packet per block in the favor of those benchmark schemes for comparison purpose.
In contrast, our proposed scheme does not require the successful reception of $M$ coded packets per batch. Therefore, the downloading rates of two benchmark schemes at the start of Phase 2 are lower than the proposed scheme. Moreover, the proposed scheme has a significant improvement in the delay in Phase 2. After disconnecting with the RSU for a certain period of time, some vehicles may leave the formed group to head for other directions. So the shorter the downloading delay is, the more reliable the V2V communication will be. Our proposed scheme performs well in enhancing the system reliability.
\subsubsection{Impact of network dynamics}
\begin{figure}
\centering
\includegraphics[width=0.47\textwidth]{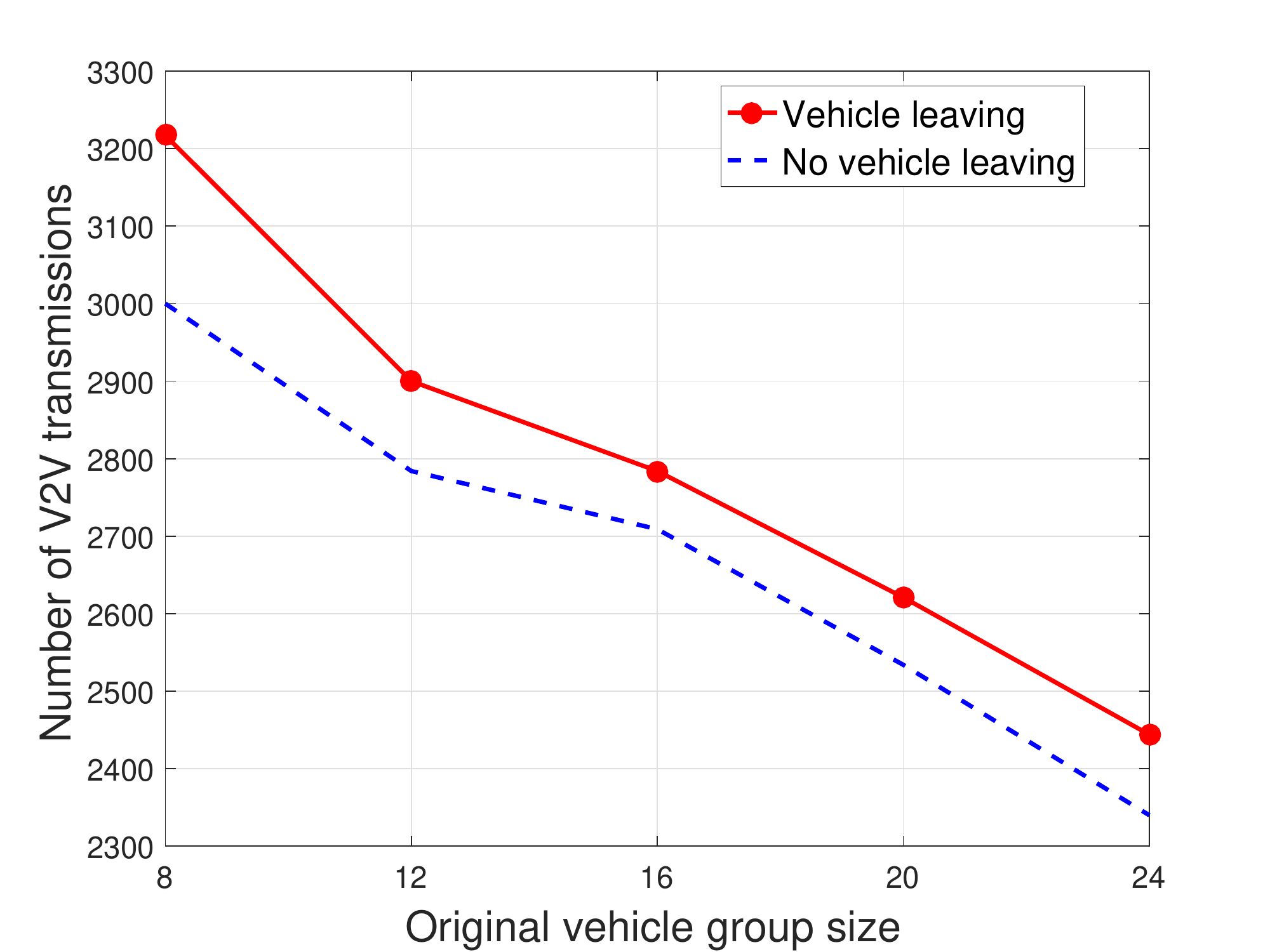}
\vspace{-0.2cm}
\caption{The impact of network dynamics on the number of P2P transmissions. Speed: $55$km/h$\pm5$km/h.}
\label{leaving}
\end{figure}

In Fig. \ref{leaving}, we assume that after passing through the communication range of the RSU, 25\% arbitrary vehicles will leave the vehicle group. Based on the simulation result shown in the figure, we can see that our proposed scheme is able to work effectively under this scenario, which needs on average 4\% increase of the number of V2V transmissions for compensation.
\section{Conclusion}\label{section_conclusion}
In this paper, we study a content distribution scenario where a RSU actively distributes a large-size file to a group of vehicles that pass by in finite time.  We propose an efficient and reliable V2X communication scheme that involves a joint I2V and V2V transmission phases, incorporating rateless-coded broadcast, network coded transmission and distributed transmission scheduling, to allow vehicles to successfully decode the file in shortest time, with least transmission overhead. Without requiring the exchange of reception status of the packets received in the first I2V broadcast phase, every vehicle distributedly computes the utility of its coded packets to prioritize the sequence of the second V2V transmission phase. The performance of the proposed scheme has been evaluated and validated by simulations. In comparison with the existing protocols such as NC-based RLS-SNR and CodeOnBasic, our proposed scheme significantly shortens the downloading delay at each vehicle. When the vehicle group size increases, by using our proposed scheme, the delay time can be reduced, while the existing protocols tend to suffer the opposite effect.
\ifCLASSOPTIONcaptionsoff
  \newpage
\fi
\bibliographystyle{IEEEtran}
\bibliography{IEEEabrv,JNL_BIB}

\begin{thebibliography}{10}
\providecommand{\url}[1]{#1}
\csname url@samestyle\endcsname
\providecommand{\newblock}{\relax}
\providecommand{\bibinfo}[2]{#2}
\providecommand{\BIBentrySTDinterwordspacing}{\spaceskip=0pt\relax}
\providecommand{\BIBentryALTinterwordstretchfactor}{4}
\providecommand{\BIBentryALTinterwordspacing}{\spaceskip=\fontdimen2\font plus
\BIBentryALTinterwordstretchfactor\fontdimen3\font minus
  \fontdimen4\font\relax}
\providecommand{\BIBforeignlanguage}[2]{{%
\expandafter\ifx\csname l@#1\endcsname\relax
\typeout{** WARNING: IEEEtran.bst: No hyphenation pattern has been}%
\typeout{** loaded for the language `#1'. Using the pattern for}%
\typeout{** the default language instead.}%
\else
\language=\csname l@#1\endcsname
\fi
#2}}
\providecommand{\BIBdecl}{\relax}
\BIBdecl

\bibitem{hartenstein2009vanet}
H.~Hartenstein and K.~Laberteaux, \emph{{VANET}: vehicular applications and
  inter-networking technologies}.\hskip 1em plus 0.5em minus 0.4em\relax John
  Wiley \& Sons, 2009, vol.~1.

\bibitem{ITS}
\BIBentryALTinterwordspacing
``{ITS} standards program.'' [Online]. Available:
  \url{https://standards.its.dot.gov/LearnAboutStandards/ResearchInitiatives}
\BIBentrySTDinterwordspacing

\bibitem{lee2002demand}
S.-J. Lee, W.~Su, and M.~Gerla, ``On-demand multicast routing protocol in
  multihop wireless mobile networks,'' \emph{Mobile networks and applications},
  vol.~7, no.~6, pp. 441--453, 2002.

\bibitem{nandan2005co}
A.~Nandan, S.~Das, G.~Pau, M.~Gerla, and M.~Sanadidi, ``Co-operative
  downloading in vehicular ad-hoc wireless networks,'' in \emph{Second Annual
  Conference on Wireless On-demand Network Systems and Services, 2005}.\hskip
  1em plus 0.5em minus 0.4em\relax IEEE, 2005, pp. 32--41.

\bibitem{lee2007first}
K.~C. Lee, S.-H. Lee, R.~Cheung, U.~Lee, and M.~Gerla, ``First experience with
  cartorrent in a real vehicular ad hoc network testbed,'' in \emph{2007 Mobile
  Networking for Vehicular Environments}.\hskip 1em plus 0.5em minus
  0.4em\relax IEEE, 2007, pp. 109--114.

\bibitem{ahlswede2000network}
R.~Ahlswede, N.~Cai, S.-Y.~R. Li, and R.~W. Yeung, ``Network information
  flow,'' \emph{IEEE Transaction on Information Theory}, vol.~46, no.~4, pp.
  1204--1216, 2000.

\bibitem{ho2006random}
T.~Ho, M.~M{\'e}dard, R.~Koetter, D.~R. Karger, M.~Effros, J.~Shi, and
  B.~Leong, ``A random linear network coding approach to multicast,''
  \emph{IEEE Transactions on Information Theory}, vol.~52, no.~10, pp.
  4413--4430, 2006.

\bibitem{lee2006code}
U.~Lee, J.-S. Park, J.~Yeh, G.~Pau, and M.~Gerla, ``Code torrent: content
  distribution using network coding in {VANET},'' in \emph{Proceedings of the
  1st international workshop on decentralized resource sharing in mobile
  computing and networking}.\hskip 1em plus 0.5em minus 0.4em\relax ACM, 2006,
  pp. 1--5.

\bibitem{ahmed2006vanetcode}
S.~Ahmed and S.~S. Kanhere, ``{VANETCODE}: network coding to enhance
  cooperative downloading in vehicular ad-hoc networks,'' in \emph{Proceedings
  of the 2006 international conference on wireless communications and mobile
  computing}.\hskip 1em plus 0.5em minus 0.4em\relax ACM, 2006, pp. 527--532.

\bibitem{firooz2012collaborative}
M.~H. Firooz and S.~Roy, ``Collaborative downloading in {VANET} using network
  coding,'' in \emph{IEEE International Conference on Communications (ICC),
  2012}.\hskip 1em plus 0.5em minus 0.4em\relax IEEE, 2012, pp. 4584--4588.

\bibitem{ye2012efficient}
F.~Ye, S.~Roy, and H.~Wang, ``Efficient data dissemination in vehicular ad hoc
  networks,'' \emph{IEEE Journal on Selected Areas in Communications}, vol.~30,
  no.~4, pp. 769--779, 2012.

\bibitem{zhu2015multiple}
W.~Zhu, D.~Li, and W.~Saad, ``Multiple vehicles collaborative data download
  protocol via network coding.'' \emph{IEEE Transactions on Vehicular
  Technology}, vol.~64, no.~4, pp. 1607--1619, 2015.

\bibitem{xing2017stochastic}
W.~Xing, F.~Liu, C.~Wang, and P.~Wang, ``Stochastic analysis of network coding
  based relay-assisted {I2V} communications in intelligent transportation
  systems,'' \emph{Wireless Communications and Mobile Computing}, vol. 2017,
  2017.

\bibitem{li2011codeon}
M.~Li, Z.~Yang, and W.~Lou, ``{CodeOn}: Cooperative popular content
  distribution for vehicular networks using symbol level network coding,''
  \emph{IEEE Journal on Selected Areas in Communications}, vol.~29, no.~1, pp.
  223--235, 2011.

\bibitem{liu2016network}
K.~Liu, J.~K.-Y. Ng, J.~Wang, V.~C. Lee, W.~Wu, and S.~H. Son,
  ``Network-coding-assisted data dissemination via cooperative
  vehicle-to-vehicle/-infrastructure communications,'' \emph{IEEE Transactions
  on Intelligent Transportation Systems}, vol.~17, no.~6, pp. 1509--1520, 2016.

\bibitem{yu2012rank}
T.-X. Yu, C.-W. Yi, and S.-L. Tsao, ``Rank-based network coding for content
  distribution in vehicular networks,'' \emph{IEEE Wireless Communications
  Letters}, vol.~1, no.~4, pp. 368--371, 2012.

\bibitem{yang2014batched}
S.~Yang and R.~W. Yeung, ``Batched sparse codes,'' \emph{IEEE Transactions on
  Information Theory}, vol.~60, no.~9, pp. 5322--5346, 2014.

\bibitem{jiang2006design}
D.~Jiang, V.~Taliwal, A.~Meier, W.~Holfelder, and R.~Herrtwich, ``Design of 5.9
  {GH}z {DSRC}-based vehicular safety communication,'' \emph{Wireless
  Communications, IEEE}, vol.~13, no.~5, pp. 36--43, 2006.

\bibitem{CV2X}
\BIBentryALTinterwordspacing
``The case for cellular {V2X} for safety and cooperative driving.'' [Online].
  Available: \url{http://5gaa.org/pdfs/5GAA-whitepaper-23-Nov-2016.pdf}
\BIBentrySTDinterwordspacing

\bibitem{zeng2017channel}
F.~Zeng, R.~Zhang, X.~Cheng, and L.~Yang, ``Channel prediction based scheduling
  for data dissemination in vanets,'' \emph{IEEE Communications Letters},
  vol.~21, no.~6, pp. 1409--1412, 2017.

\bibitem{ng2013finite}
T.-C. Ng and S.~Yang, ``Finite-length analysis of {BATS} codes,'' in \emph{2013
  International Symposium on Network Coding (NetCod)}.\hskip 1em plus 0.5em
  minus 0.4em\relax IEEE, 2013, pp. 1--6.

\bibitem{zhang2015novel}
R.~Zhang, X.~Cheng, L.~Yang, X.~Shen, B.~Jiao \emph{et~al.}, ``A novel
  centralized {TDMA}-based scheduling protocol for vehicular networks.''
  \emph{IEEE Transaction on Intelligent Transportation Systems}, vol.~16,
  no.~1, pp. 411--416, 2015.

\bibitem{xu2016two}
X.~Xu, M.~S. G.~P. Kumar, Y.~L. Guan, and P.~H.~J. Chong, ``Two-phase
  cooperative broadcasting based on batched network code,'' \emph{IEEE
  Transactions on Communications}, vol.~64, no.~2, pp. 706--714, 2016.

\bibitem{guo2017enabling}
T.~Guo, C.~Li, W.~Dong, Z.~Miao, and X.~Su, ``Enabling efficient content
  dissemination for cooperative vehicular networks,'' in \emph{2017 IEEE 28th
  Annual International Symposium on Personal, Indoor, and Mobile Radio
  Communications (PIMRC)}.\hskip 1em plus 0.5em minus 0.4em\relax IEEE, 2017,
  pp. 1--5.

\bibitem{rossi2017stable}
G.~V. Rossi, Z.~Fan, W.~H. Chin, and K.~K. Leung, ``Stable clustering for
  ad-hoc vehicle networking,'' in \emph{2017 IEEE Wireless Communications and
  Networking Conference (WCNC)}.\hskip 1em plus 0.5em minus 0.4em\relax IEEE,
  2017, pp. 1--6.

\bibitem{xu2017quasi}
X.~Xu, Y.~L. Guan, Y.~Zeng, and C.-C. Chui, ``Quasi-universal {BATS} code,''
  \emph{IEEE Transactions on Vehicular Technology}, vol.~66, no.~4, pp.
  3497--3501, 2017.

\bibitem{cheng2007mobile}
L.~Cheng, B.~E. Henty, D.~D. Stancil, F.~Bai, and P.~Mudalige, ``Mobile
  vehicle-to-vehicle narrow-band channel measurement and characterization of
  the 5.9 {GH}z dedicated short range communication ({DSRC}) frequency band,''
  \emph{IEEE Journal on Selected Areas in Communications}, vol.~25, no.~8, pp.
  1501--1516, 2007.

\bibitem{wang2012joint}
Q.~Wang, P.~Fan, and K.~B. Letaief, ``On the joint {V2I} and {V2V} scheduling
  for cooperative vanets with network coding,'' \emph{IEEE Transactions on
  Vehicular Technology}, vol.~61, no.~1, pp. 62--73, 2012.

\end{thebibliography}
\end{document}